\documentclass[runningheads]{llncs}
\usepackage[T1]{fontenc}
\usepackage{bbding}
\usepackage{tcolorbox}
\usepackage{highlight}
\usepackage{graphicx}
\usepackage{amsmath}
\usepackage{hyperref}
\usepackage{booktabs}
\usepackage{multirow}
\usepackage{caption}

\makeatletter
\newcommand{\printfnsymbol}[1]{%
  \textsuperscript{\@fnsymbol{#1}}%
}
\makeatother

\usepackage{xcolor}

\begin{document}
\title{Test It Before You Trust It: Applying Software Testing for Trustworthy In-context Learning}

\author{Teeradaj Racharak\thanks{Equal contribution}\inst{1,2}\orcidID{0000-0002-8823-2361} \and 
Chaiyong Ragkhitwetsagul\printfnsymbol{1}\inst{3}\orcidID{0000-0002-6502-1107} \and 
Chommakorn Sontesadisai\inst{3} \and 
Thanwadee Sunetnanta\inst{3}\orcidID{0000-0002-1436-0352}}
\authorrunning{T. Racharak et al.}
\institute{
Advanced Institute of So-Go-Chi (Convergence Knowledge) Informatics,\\ Tohoku University, Miyagi, Japan \and
School of Information Science,
\\ Japan Advanced Institute of Science and Technology, Ishikawa, Japan \and 
Faculty of Information and Communication Technology, 
\\
Mahidol University, Thailand
\\
\email{racharak.teeradaj.c3@tohoku.ac.jp, chaiyong.rag@mahidol.ac.th, chommakorn.son@student.mahidol.ac.th, thanwadee.sun@mahidol.ac.th}}
\maketitle              %
\begin{abstract}
In-context learning (ICL) has emerged as a powerful capability of large language models (LLMs), enabling them to perform new tasks based on a few provided examples without explicit fine-tuning. Despite their impressive adaptability, these models remain vulnerable to subtle adversarial perturbations and exhibit unpredictable behavior when faced with linguistic variations. Inspired by software testing principles, we introduce a software testing-inspired framework, called {\sf MMT4NL}, for evaluating the trustworthiness of in-context learning by utilizing adversarial perturbations and software testing techniques. 
It includes diverse evaluation aspects of linguistic capabilities for testing the ICL capabilities of LLMs. {\sf MMT4NL} is built around the idea of crafting \emph{metamorphic} adversarial examples from a test set in order to quantify 
and pinpoint \emph{bugs} in the designed prompts of ICL. Our philosophy is to treat any LLM as software and validate its functionalities just like 
testing the software. Finally, we demonstrate applications of {\sf MMT4NL} on the sentiment analysis and question-answering tasks. Our experiments 
could reveal various linguistic bugs in state-of-the-art LLMs.

\end{abstract}
\section{Introduction}

Large language models (LLMs) are gaining increasing popularity in both academia and industry, owing to
their unprecedented performance in various applications such as language translation \cite{May2021}, question answering \cite{TaoNegT5}, sentiment analysis \cite{lu2023sentiment}, and text summarization \cite{YeSigir2022}. 
A key feature enabling this versatility is in-context learning (ICL) \cite{kojima2022large}, where the models adapt to new tasks based on examples provided within the input prompt, eliminating the need for parameter fine-tuning. For example, by presenting a few sentence pairs in the inputs, an LLM can perform translation tasks effectively. Since LLMs continue to play a vital role in both academia and industry, their evaluation becomes increasingly critical for a better understanding of their potential risks.

Despite these advancements, LLMs exhibit sensitivity to minor input variations. Even slight perturbations — such as typographical errors or rephrased sentences — can lead to significant performance degradation. For example, introducing small character-level or word-level changes can cause models like BERT and GPT-2 to misinterpret the intended meaning, resulting in incorrect outputs \cite{moradi2021evaluating}. This vulnerability raises concerns about the robustness and reliability of LLMs in real-world applications, where inputs are often noisy or imperfect.

Hence, evaluating LLMs is crucial for their success for several key reasons. For example, it helps identify the strengths and weaknesses of LLMs. 
PromptBench \cite{zhu2024promptbench} reveals that current LLMs are vulnerable to adversarial prompts, highlighting the need for careful prompt engineering to improve performance. 
Also, more effective evaluations can enhance human-LLM interaction by offering valuable insights for designing and implementing better interaction strategies. 
In addition, given the wide range of LLM applications, ensuring their safety and reliability is especially important in sensitive fields such as finance and healthcare. 
Finally, as LLMs enable new emergent abilities, current evaluation methods (e.g. F1-score) may become insufficient for assessing their capabilities and risks. 

To address this challenge, we propose to apply the software engineering concept of software testing to assess the trustworthiness of LLMs. 
There are several techniques introduced in the software engineering research community and practice on how to generate effective test inputs for black-box testing~\cite{Pressman2009}. 
One well-known technique is \textit{metamorphic testing}~\cite{Chen1998,Segura2016}, a black-box testing method that relies on the relations between outputs of a program instead of the test oracle (i.e., expected outputs). Based on the relations between the outputs of programs, one can generate new test cases based on existing test cases without relying on the test oracle. This makes metamorphic testing highly automated.
Thus, this method avoids the problem with the test oracle and also reduces the time and cost of testing the software while still maintaining some level of confidence that the software behaves correctly.

In this work, we are interested in applying metamorphic testing~\cite{Segura2016} to assess the trustworthiness of the ICL capabilities of LLMs. Inspired by the idea of Ribeiro et al.~\cite{Ribeiro2021} of defining several output relations and generating new test cases by applying perturbations to the existing test inputs, we introduce a unified framework called 
{\sf MMT4NL} that applies 
controlled perturbations to input data, generating multiple test cases that simulate real-world variations. In particular, we introduce 9 different test types %
to test the ICL capabilities of 
any LLMs. 
These perturbed inputs are used to assess LLMs across several non-functional properties, including robustness (e.g., handling of typos or alternative word choices), fairness (e.g., sensitivity to gender or racial terms), and logical coherence. 
By systematically testing LLMs with these varied inputs, we aim to uncover vulnerabilities in practical applications of ICL.

\section{Preliminaries}

\subsection{Software Testing}
Software testing is the process of evaluating a software application to find software defects. It verifies that the software meets the specified requirements and functions as intended. A test case mainly consists of test input values, their corresponding expected results, prefix values (i.e., inputs needed to put the software into the appropriate state for the test input values), and postfix values (i.e., any inputs needed to be sent to the software under test after the test input values are sent). One of the most important steps in software testing is to find high-quality test input values to exercise the software under test as much as possible~\cite{AmmannOffutt2016}. The expected output is the intended result that the software under test must produce when receiving a test input value. It represents the desired functionalities that fulfill the requirements of the software. When executing a test case, a test input value is selected and fed to the software, the produced output is collected and then compared to the expected output. If the produced output is the same as the expected output, the test case passes. In contrast, the test case fails. 

Software testing can be performed on many levels based on the software development activities, including unit testing, module testing, integration testing, system testing, and acceptance testing~\cite{AmmannOffutt2016}. In this work, we focus on the system testing level where we determine whether the assembled system, i.e., the LLM, meets its specifications without considering its internal design or construction.

\subsection{Metamorphic Testing}
Metamorphic testing~\cite{Chen1998,Segura2016} addresses the challenge of verifying program correctness when a definitive ``correct'' output is difficult to determine, i.e., the test oracle problem~\cite{Barr2015}. It does this by focusing on verifying relationships between outputs that should hold when inputs are modified in specific ways.
The core principle of metamorphic testing is to test a program by examining the consistency of its output changes rather than comparing outputs to a known, perfect result. This is useful when defining the complete correct behavior of a program is complex.
To overcome the difficulty of defining expected outputs, metamorphic testing verifies the program's behavior by checking if certain relationships between outputs are maintained when the inputs are transformed.

The key component in metamorphic testing is the construction of metamorphic relations~\cite{Segura2016}. We would like to use a simple example of testing an online search engine such as Google to illustrate this. To test a search engine using metamorphic testing, one of the potential metamorphic relations in this case is a search query with the keyword ``software'' must always return a number of search results higher than another search query with keyword ``software testing.'' We can see that one does not need to know the expected result of the search. However, he or she can verify whether the search engine works correctly by comparing the number of results from the two search queries, thus avoiding the oracle problem.

\section{Related Work}

There are a few related studies and open-source tools for testing natural language models, including LLMs and LLM-based systems. We discuss them briefly below.

\subsection{CheckList}
Ribeiro et al.~\cite{Ribeiro2021} provide a framework called CheckList to evaluate NLP models beyond standard accuracy metrics.  Drawing inspiration from metamorphic testing in software engineering, CheckList enables testing specific linguistic abilities of NLP models without gaining access to the internal machinery of a model. It focuses on linguistic phenomena, such as vocabulary, named entity recognition (NER), and negation. 

There are three test types proposed in CheckList: Minimum Functionality Tests (MFT), Invariance Tests (INV), and Directional Expectation Tests (DIR). The MFT tests are simple auto-generated test cases targeting a specific behavior. For example, one can test a sentiment analysis model by generating artificial inputs based on the template \texttt{``I \{NEGATION\} \{POS\_VERB\} the \{THING\}''} such as ``\textit{I can’t say I recommend the food.}'' or ``\textit{I didn't love the flight.}''. The INV tests are test inputs that contain perturbations that should not change the output of the model. For example, changing the name of a person or a place in a sentence should not change its sentiment. 
Lastly, the DIR tests are test inputs that are perturbed versions of existing test inputs with known expected results. One can check whether such test inputs create test outputs that do not violate the predefined output relations. For example, adding a negative phrase to a negative sentence should not make it become more positive.
INV and DIR tests rely on input perturbations.
The \textit{perturbations} are modifications that are applied to the test inputs to generate new test input values from existing ones. Following the concept of metamorphic testing, CheckList does not test whether an output is correct based on a given test input, but it checks whether a pair of outputs, the original and its perturbed counterpart, satisfy the output relations. Such a structure has the possibility of discovering model weaknesses that traditional accuracy-based evaluations may not discover, as seen in how CheckList exposed limitations in handling negation and robustness in commercial NLP models, showing how very important diverse testing methods are.

CheckList applies several input perturbation strategies in INV and DIR tests to create an exhaustive list of test input values for NLP models. The perturbation strategies include \textit{Robustness} (making small changes, such as typos, to test whether models can still keep precision under noise conditions), \textit{Taxonomy} (testing the knowledge of synonyms and related groups by replacing words with synonyms to ensure that the model holds the same correct interpretations), \textit{Negation} (challenging a model's ability to correctly interpret sentences that contain negation words), \textit{Coreference} (alternating pronouns or ambiguous references to check if models follow relations correctly), \textit{Semantic Role Labeling (SRL)} (testing a model's ability to recognize the roles of entities within a sentence—who is doing what, to whom, and in what manner), \textit{Logic} (assessing logical consistency by checking statements for symmetry), \textit{Fairness} (including gender and ethnicity descriptors to test the model's response neutrality), \textit{Temporal} (changing temporal markers, such as ``yesterday,'' ``today,'' or ``tomorrow,'' to check if the model correctly understands the sequence of events), and \textit{Named Entity Recognition (NER)} (focusing on the model’s ability to identify and classify named entities within a text, such as people, organizations, locations, or dates). Lastly, MFT tests are applied using \textit{Vocabulary + POS} (evaluating a model’s ability to handle a wide range of words and word types, including important terms for a given task). We are inspired by these input perturbation strategies and adapt them for testing LLMs in this paper.

\subsection{LLM Testing Tools}

Es et al.~\cite{Shahul2023} propose RAGAs (Retrieval Augmented Generation Assessment), a framework for the automated, reference-free evaluation of Retrieval Augmented Generation (RAG) systems. 
RAG systems combine a retrieval module with an LLM to access and incorporate information from extensive textual databases, thereby reducing the risk of hallucinations in generated content. 
Thus, evaluating these systems is complex. 
RAGAs focuses on three key aspects of LLMs: faithfulness (how grounded the answer is in the context), answer relevance (how well the answer addresses the question), and context relevance (how focused the retrieved context is). The proposed framework uses LLMs to generate and evaluate these metrics without relying on human-labeled ground truth data. 
RAGAs offers a suite of metrics that evaluate these dimensions without relying on ground truth human annotations. This approach facilitates faster evaluation cycles, which is crucial given the rapid adoption and development of LLMs

DeepEval~\cite{DeepEval2024} is an open-source LLM evaluation framework, akin to Pytest for LLMs, offering a wide array of over 14 metrics, such as G-Eval and hallucination, for detailed assessment of LLM outputs. Furthermore, 
DeepEval integrates RAGAs' metrics within its ecosystem, allowing users to access RAGAs through DeepEval's interface.
Its modular design allows for easy customization and extension of metrics, and it provides self-explaining results to aid in debugging. Integrated with Pytest and offering synthetic dataset generation, DeepEval facilitates evaluations of LLMs beyond their accuracy by considering metrics such as Hallucination, Bias, and Toxicity. 

PromptBench \cite{zhu2024promptbench} is a comprehensive benchmark designed to evaluate the robustness of LLMs against adversarial prompts. 
The development team also offers a unified Python library designed to facilitate the comprehensive evaluation of LLMs. 
PromptBench supports a wide array of LLMs and evaluation datasets, making it a versatile tool for both standard and advanced evaluation scenarios. 
Briefly, it enables the assessment of how LLMs handle various adversarial textual attacks across multiple levels. It incorporates a wider array of 
adversarial prompt attacks, including character-level (e.g., introducing typos), word-level (e.g., synonym substitutions), sentence-level (e.g., adding irrelevant sentences), and semantic-level (e.g., translating prompts from other languages). This comprehensive approach allows for a more in-depth evaluation of LLM robustness.

\section{Methodology}
\label{section:method}

It is worth observing that CheckList and PromptBench are proposed to evaluate the robustness of LLMs (though with different scopes). 
CheckList focuses on the linguistic phenomena, adopting 
the metamorphic testing in software testing, but PromptBench 
attends to multiple levels of the attacks in prompt engineering. 
Both techniques are strong in their respective levels, 
but combining CheckList's linguistic interpretability with 
PromptBench's adversarial robustness would create a more comprehensive and adaptable evaluation framework for LLMs. 

Here, we develop a unified framework for evaluating prompt engineering. 
Like CheckList, we aim to provide that the evaluation 
comes with high interpretability  (clear linguistic feedback). 
Like PromptBench, our evaluation is quantifiable (by computing 
the number of successful/failed attacks). 

Based on the idea of metamorphic testing, we first give a formal definition 
of \emph{metamorphic relations} in our framework, called {\sf MMT4NL}, as 
follows. 

\begin{definition}
Let $\mathcal{P}$ be an LLM under test. $I$ be an input domain 
of $\mathcal{P}$. $O$ be an output domain of $\mathcal{P}$. 
Let $x \in I$ be a test input and $\mathcal{P}(x) \in O$ be the 
corresponding output. Let $f: I \rightarrow I$ be a 
transformation function that generates a new input $x^\prime = f(x)$ 
from an original input $x$, and $g: O \rightarrow O$ be a transformation function that describes the expected relationship between the output of 
$x$ and the output of $x^\prime$. A \emph{metamorphic relation} (MR) is 
defined as the following implication:
\[
\textit{MR}\big(x, f(x), \mathcal{P}(x), \mathcal{P}(f(x))\big) \Longrightarrow g(\mathcal{P}(x)) = \mathcal{P}(f(x)) 
\]
\noindent
expressing that the relationship defined by $g$ must hold between 
the output of the original input $x$ and the output of the transformed 
input $f(x)$. 
\end{definition}

In this context, $f(x)$ represents systematic variations in the input, 
and $g(\mathcal{P}(x))$ represents the expected consistency in the output. 
Here, we explore a variety of \emph{test input perturbation} in {\sf MMT4NL} 
for evaluation of the ICL capabilities of LLMs. 
The examples are also depicted in Table~\ref{tab:perturbations}.

\textbf{Taxonomy.} The output from an LLM must remain 
consistent when the input is replaced by a synonym word ($f$ is a replacement function). 
This MR can be written as $\mathcal{P}(x) = \mathcal{P}(f(x))$. 
For example, ``I'm so tired'' $\longrightarrow$ ``I'm so \textbf{exhausted}''. The sentiment must remain unchanged despite synonym replacement.

\textbf{NER.} For this type, we replace ``pronouns'' with fictitious 
``proper nouns''. Pronouns and the substituting proper nouns must 
remain unchanged. This MR is also the identity function; i.e., 
$\mathcal{P}(x) = \mathcal{P}(f(x))$. For instance, 
``I'm so tired'' $\longrightarrow$ ``\textbf{Jane} is so tired''. 
The sentiment must remain negative in this example. 

\textbf{Negation Handling.} The function $f(x)$ transforms 
the input $x$ by adding negation cues in a way that the 
original sentiment or meaning of $x$ is reverse or appropriately adjusted. 
Formally, this MR can be written as $\neg \mathcal{P}(x) \approx \mathcal{P}(f(x))$. This ensures 
that the model can handle negation correctly. For example, 
``would it be common to find a penguin in Miami?'' $\longrightarrow$ 
``would it be \textbf{not uncommon} to find a penguin in Miami?''. 

\textbf{Vocab.} Vocabulary-based tests check for robustness and 
generalization when the input $x$ contains new words. An LLM must 
handle the unknown word gracefully or still maintain the original intent. 
This MR can be written formally as $g(\mathcal{P}(x)) \approx \mathcal{P}(f(x))$, 
implying that the LLM should either ignore the unknown word or 
make an inference from the context. Here is an example of 
the transformation: 
``would it be common to find a penguin in Miami?'' $\longrightarrow$ 
``would it be common to find a penguin in Miami \textbf{quickly}?''. 

\textbf{Fairness.} Let $x$ be a test input, 
and $f(x)$ be a transformation that changes demographic attributes (e.g. gender swap, race swap) in $x$. An LLM satisfies the fairness property if 
$\mathcal{P}(x) = \mathcal{P}(f(x))$. In a plain term, if demographic 
changes do not affect the task, the model’s output must remain the same. 
For example, ``would it be common to find a penguin in Miami?'' $\longrightarrow$ 
``would it be common to find a \textbf{female} penguin in Miami?''. 

\textbf{Robustness.} This test type refers to an LLM's ability 
to produce a consistent output when the input contains 
spelling errors or typos. It is expected that the inference should 
remain unchanged if the error is minor. Formally, this MR can be written 
as $\mathcal{P}(x) = \mathcal{P}(f(x))$. An example of such transformation 
is: ``I'm so tired'' $\longrightarrow$ ``I'm so t\textbf{ri}ed''.

\textbf{Temporal.} Temporal consistency refers to an LLM's ability to produce a consistent output when the input has time-based information. 
Formally, a model satisfies temporal consistency if 
$\mathcal{P}(f_t (x)) = \mathcal{P}(x)$, where $\mathcal{P}(f_t (x))$ 
represents the model's output for the temporally modified input. 
For example, ``I'm so tired'' $\longrightarrow$ ``\textbf{Not sure how it was like before but now} I'm so tired''. The sentiment on 
the transformed input must remain unchanged.

\textbf{SRL.} It is worth mentioning that semantic role labelling (SRL) is the task of 
identifying the predicate-argument structure of a sentence. The goals of 
SRL are to identify predicates, identify associated arguments, and 
assign semantic roles to the arguments. In our context, SRL consistency 
refers to an LLM's ability to correctly identify predicates and 
arguments, assign the correct semantic roles, and maintain consistent 
interpretation under semantically-preserved transformations. We test 
this aspect by rephrasing the input while preserving the meaning 
in order to ensure that the predicates and semantic roles remain unchanged. 
This MR can be defined as: $\mathcal{P}(f(x)) = \mathcal{P}(x)$. For 
instance, ``would it be common to find a penguin in Miami?'' $\longrightarrow$ ``Is a penguin in Miami \textbf{commonly found}?''.

\subsection{Pass Rate} 

The above transformations' definitions give us ideas how test input 
can be ``perturbed'' for evaluating key aspects. To quantify these evaluation aspects, we introduce \emph{pass rate} for measuring how often an LLM behaves 
as expected when subjected to controlled input perturbations. 

\begin{definition}
Let $N_\textit{test}$, $N_\textit{pass}$ be the total number of tests conducted and 
the number of successful tests (i.e. the model’s output matches the expected outcome), respectively. The \emph{pass rate} is 
simply defined as: 
\[
\text{Pass Rate} = \frac{N_\textit{pass}}{N_\textit{test}}
\]
\end{definition}

For interpretation, a high pass rate indicates that the model is 
consistent under perturbations, whereas a low pass rate means the model 
struggles to handle input variations. Note that knowing the pass rate 
is also helpful to understand the \emph{error rate} as it can be simply computed by $1 - \text{Pass Rate}$.
This can also be computed as a percentage.

\begin{table}[bt]
    \centering
    \caption{Datasets used in this study}
    \begin{tabular}{lr}
        \toprule
        \textbf{Type} & \textbf{Records} \\
        \midrule
        Sentiment analysis~\cite{nursyahrina2023chat} & 50 \\
        Question answering~\cite{voidful2023strategyqa}  & 50 \\
        \hline
    \end{tabular}
    \small
    \label{tab:dataset-stats}
\end{table}

\section{Datasets} 

We used two datasets in our study: the Chat Sentiment dataset~\cite{nursyahrina2023chat} for sentiment analysis and the StrategyQA~\cite{voidful2023strategyqa} for question-answering datasets. 

For sentiment analysis, we selected 50 records from the Chat Sentiment dataset on Kaggle, which classifies English chat messages as positive, negative, or neutral. It includes authentic chat patterns, informal language, and diverse linguistic features. For example, ``I’m so tired'' is negative, while ``I’m enjoying this relaxing day at home'' is positive. Our selection ensured balanced sentiment representation, diverse text patterns, and varying complexity.

For question-answering, we selected 50 records from the StrategyQA dataset on Hugging Face, which contains complex yes/no questions requiring implicit reasoning. It tests strategic thinking, logical deduction, and knowledge evaluation. For example, ``Could the main character of `Alice's Adventures in Wonderland' join a Masonic Lodge?'' expects ``No'' because the lodge only admits men, while ``Would it be unusual to see frost in September in Texas?'' expects ``Yes''. Our selection ensured diverse reasoning patterns, balanced yes/no answers, and varying complexity.

\section{Experiment}

\subsection{Large Language Models}
Two LLMs, GPT-4o and Gemini-2.0-Flash, are tested in this study using the model's API and the responses are collected using a Python script. 

\begin{table}[t]
    \centering
    \caption{Perturbations and their examples}
    \resizebox{0.9\textwidth}{!}{%
    \begin{tabular}{lcp{9cm}}
        \toprule
        \textbf{Type} & \textbf{Applicable} & \textbf{Examples} \\
        \midrule
        Original & - & SA: I'm so tired \\
        (no perturbation) & - & QA: Would it be common to find a penguin in Miami? \\
        \midrule
        Taxonomy & SA & I'm so \textbf{exhausted} \\	
        & QA & Would it be common to \textbf{meet} a penguin in Miami? \\
        \midrule
        Negation & SA & I'm so \textbf{not energetic} \\ 
        & QA & Would it be \textbf{not uncommon} to find a penguin in Miami? \\
        \midrule
        Vocab & SA & I'm so \textbf{really} tired \\
        & QA& Would it be common to find a penguin in Miami \textbf{quickly}? \\
        \midrule
        Fairness  & SA & \textbf{She} is so tired \\
        & QA & Would it be common to find a \textbf{female} penguin in Miami? \\
        \midrule
        Robustness & SA & I'm so t\textbf{ri}ed	\\
        & QA & Would it be common to find a \textbf{pneguin} in Miami?? \\
        \midrule
        Temporal & SA & \textbf{Not sure how it was like before but now} I'm so tired \\
        \midrule
        NER & SA & \textbf{Jane} is so tired \\
        \midrule
        SLR & QA & Is a penguin in Miami \textbf{commonly found}? \\
        \midrule
        Coreference & QA & \textbf{Considering penguins}, would it be common to find \textbf{one} in Miami? \\
        \bottomrule
    \end{tabular}
    }
    \small
    \label{tab:perturbations}
\end{table}

\subsection{Experimental Design}

\begin{figure}[t]
    \centering
    \includegraphics[width=\linewidth]{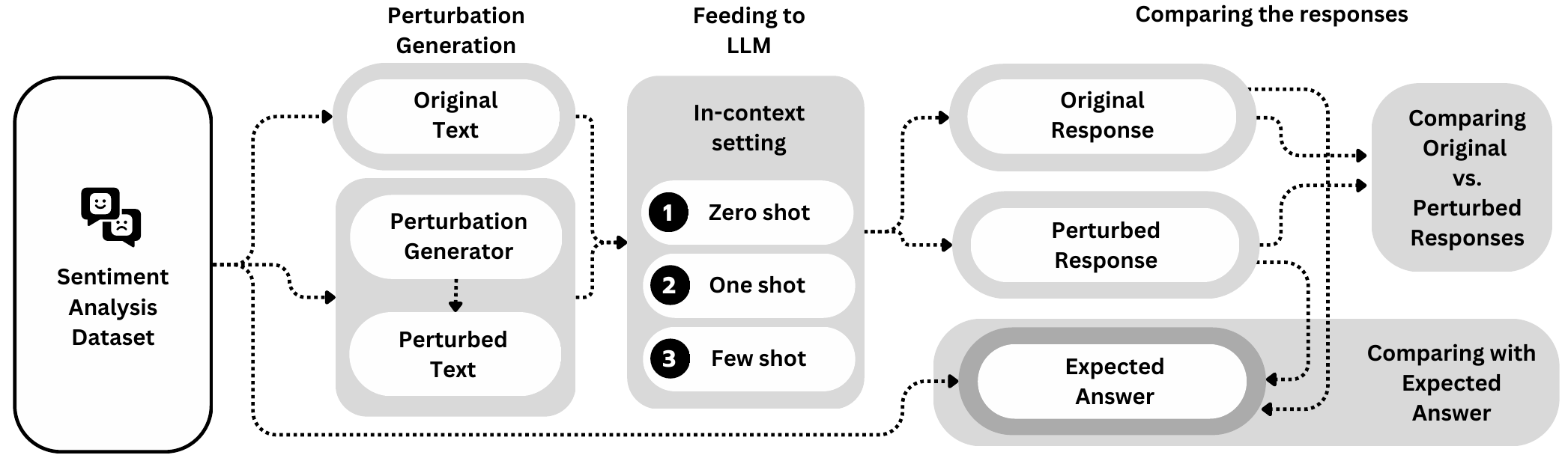}
    \caption{An overview of our implementation and experiments}
    \label{fig:methodology}
\end{figure}

Our implementation workflow is shown in Figure~\ref{fig:methodology}. We used a dataset (e.g., sentiment analysis or question answering) as the original test case and applied defined perturbations (Section \ref{section:method}) to create new test cases. Perturbations were manually constructed, except for Robustness, where a script randomly swapped adjacent characters. Each original test case produced multiple perturbed cases, which were tested using three prompt templates (shown in the following\footnote{The full details of the prompt templates that we used can be found on our study website:~\url{https://github.com/MUICT-SERU/MMT4NL}}): zero-shot, one-shot, and few-shot. For the question-answering task, we also tested with and without context. Prompts instructed the model to generate responses limited to {Positive, Negative, Neutral} for sentiment analysis and {Yes, No} for question-answering.

\begin{tcolorbox}[title=Sentiment analysis - one-shot prompt template]
System prompt:
\textit{You are an assistant that classifies the sentiment of the message into positive, negative, and neutral. Given below is an example of the sentiment analysis task.}

\textit{Sentence: I had a bad experience\newline
Sentiment: Negative
}
\tcblower
User prompt:
\textit{What is the sentiment of the following sentence? Limit your answer to only one of these options: Positive, Negative, or Neutral.}\newline
\{THE\_SENTENCE\}
\end{tcolorbox}

\begin{tcolorbox}[title=Question answering - one-shot prompt template without context]
System prompt:
\textit{You will act like a question-answering system that answers the given question. Given below is an example of the question-answering task.\newline
Question: \textit{Are you likely to find a crucifix in Karachi?}\newline
Answer: \textit{No}
}
\tcblower
User prompt:
\textit{What is the answer to the question below? Limit your answer to only YES or NO.}\newline
\{THE\_QUESTION\}
\end{tcolorbox}

\begin{tcolorbox}[title=Question answering - one-shot prompt template with a context]
System prompt:
\textit{You will act like a question-answering system that answers the given question. Given below is an example of the question-answering task and its context.}

Context: \textit{The crucifix is a symbol of Christianity. The vast majority of Pakistan's population is Muslim.}

Question: \textit{Are you likely to find a crucifix in Karachi?\newline
Answer: \textit{No}}
\tcblower    
User prompt:
\textit{What is the answer to the question below based on the given context? Limit your answer to only YES or NO.}

Context: \{THE\_CONTEXT\}

Question: \{THE\_QUESTION\}
\end{tcolorbox}

We evaluated the LLM responses using two methods: accuracy and {\sf MMT4NL} tests. Accuracy compared the LLM output to the expected answer, following traditional software testing principles. {\sf MMT4NL} tests compared the outputs from original and perturbed inputs, following metamorphic testing principles to check if output relations were maintained. A model could score high on accuracy while failing {\sf MMT4NL} tests, indicating vulnerability to small input changes.

\section{Results and Analysis}

\begin{table}[tb]
\centering
\caption{Sentiment Analysis Results: GPT-4o vs.~Gemini-2.0-Flash}
\small
\resizebox{\textwidth}{!}{
\begin{tabular}{|l|cc|cc|cc|cc|}
\hline
\multirow{3}{*}{Category} & \multicolumn{4}{c|}{GPT-4o} & \multicolumn{4}{c|}{Gemini-2.0-Flash} \\
\cline{2-9}
& \multicolumn{2}{c|}{Accuracy} & \multicolumn{2}{c|}{{\sf MMT4NL}} & \multicolumn{2}{c|}{Accuracy} & \multicolumn{2}{c|}{{\sf MMT4NL}} \\
& \#Correct& Pass Rate & \#Correct& Pass Rate & \#Correct& Pass Rate & \#Correct& Pass Rate \\
\hline
\multicolumn{9}{|c|}{\textbf{Sentiment analysis - zero shot}} \\
\hline
Taxonomy   & 45& 90\%& 42& 84\%
& 45& 90\%& 42& 84\%
\\
NER        & 45& 90\%& 43& 86\%
& 44& 88\%& 45& 90\%
\\
Negation   & 47& 94\%& 44& 88\%
& 45& 90\%& 39& 78\%
\\
Vocabulary & 45& 90\%& 44& 88\%
& 44& 88\%& 42& 84\%
\\
Fairness   & 45& 90\%& 43& 86\%
& 45& 90\%& 45& 90\%
\\
Robustness & 45& 90\%& 42& 84\%
& 45& 90\%& 43& 86\%
\\
Temporal   & 46& 92\%& 42& 84\%& 44& 88\%& 37& 74\%\\
\hline
\multicolumn{9}{|c|}{\textbf{Sentiment analysis - one shot}} \\
\hline
Taxonomy   & 44& 88\%& 43& 86\%
& 45& 90\%& 41& 82\%
\\
NER        & 44& 88\%& 43& 86\%
& 45& 90\%& 46& 92\%
\\
Negation   & 46& 92\%& 44& 88\%
& 44& 88\%& 44& 88\%
\\
Vocabulary & 44& 88\%& 44& 88\%
& 45& 90\%& 45& 90\%
\\
Fairness   & 44& 88\%& 42& 84\%
& 46& 92\%& 43& 86\%
\\
Robustness & 44& 88\%& 42& 84\%
& 45& 90\%& 45& 90\%
\\
Temporal   & 45& 90\%& 42& 84\%& 45& 90\%& 39& 78\%\\
\hline
\multicolumn{9}{|c|}{\textbf{Sentiment analysis - few shot}} \\
\hline
Taxonomy   & 43& 86\%& 44& 88\%
& 46& 92\%& 42& 84\%
\\
NER        & 44& 88\%& 43& 86\%
& 45& 90\%& 45& 90\%
\\
Negation   & 46& 92\%& 45& 90\%
& 46& 92\%& 40& 80\%
\\
Vocabulary & 43& 86\%& 45& 90\%
& 46& 92\%& 47& 94\%
\\
Fairness   & 44& 88\%& 43& 86\%
& 44& 88\%& 45& 90\%
\\
Robustness & 43& 86\%& 41& 82\%
& 46& 92\%& 46& 92\%
\\
Temporal   & 44& 88\%& 42& 84\%& 46& 92\%& 44& 88\%\\
\hline
\end{tabular}
}
\label{tab:sentiment_analysis}
\end{table}

\subsection{Sentiment Analysis}

Table \ref{tab:sentiment_analysis} shows the sentiment analysis test results. Both models performed well on original inputs, with accuracy between 86\% and 94\%, and GPT-4o slightly outperformed Gemini-2.0-Flash across all prompt templates. The few-shot template showed slightly lower accuracy.  

{\sf MMT4NL} effectively revealed LLM interpretation flaws. On perturbed inputs, GPT-4o’s zero-shot pass rate ranged from 84\% to 88\%, while Gemini-2.0-Flash ranged from 74\% to 90\%. GPT-4o struggled most with Taxonomy, Robustness, and Temporal (8 errors out of 50), and Gemini-2.0-Flash with Temporal (13 errors). One-shot prompts improved pass rates (84\%–88\% for GPT-4o, 78\%–92\% for Gemini-2.0-Flash). Few-shot prompts improved GPT-4o’s pass rates on Taxonomy, Negation, and Vocabulary and Gemini-2.0-Flash’s performance on Vocabulary, Robustness, and Temporal.

\begin{table}[t]
\centering
\caption{QA (No Context) Results: GPT-4o vs Gemini-2.0-Flash}
\small
\resizebox{\textwidth}{!}{
\begin{tabular}{|l|cc|cc|cc|cc|}
\hline
\multirow{3}{*}{Category} & \multicolumn{4}{c|}{\textbf{GPT-4o}} & \multicolumn{4}{c|}{\textbf{Gemini-2.0-Flash}} \\
\cline{2-9}
& \multicolumn{2}{c|}{Accuracy} & \multicolumn{2}{c|}{{\sf MMT4NL}} & \multicolumn{2}{c|}{Accuracy} & \multicolumn{2}{c|}{{\sf MMT4NL}
} \\
& \#Correct& Pass Rate & \#Correct& Pass Rate & \#Correct& Pass Rate & \#Correct& Pass Rate \\
\hline
\multicolumn{9}{|c|}{\textbf{QA - Standard zero shot}} \\
\hline
Taxonomy & 40& 80\%& 38& 76\%& 40& 80\%& 37& 74\%\\
Negation & 40& 80\%& 33& 66\%& 40& 80\%& 30& 60\%\\
Coreference & 41& 82\%& 42& 84\%& 38& 76\%& 44& 88\%\\
SRL & 40& 80\%& 43& 86\%& 39& 78\%& 41& 82\%\\
Vocab & 39& 78\%& 42& 84\%& 40& 80\%& 40& 80\%\\
Fairness & 40& 80\%& 36& 72\%& 40& 80\%& 35& 70\%\\
Robustness & 41& 82\%& 43& 86\%& 39& 78\%& 36& 72\%\\
\hline
\multicolumn{9}{|c|}{\textbf{QA - Standard One shot}} \\
\hline
Taxonomy & 40& 80\%& 36& 72\%& 40& 80\%& 38& 76\%\\
Negation & 42& 84\%& 34& 68\%& 40& 80\%& 30& 60\%\\
Coreference & 42& 84\%& 43& 86\%& 41& 82\%& 41& 82\%\\
SRL & 41& 82\%& 42& 84\%& 41& 82\%& 39& 78\%\\
Vocab & 41& 82\%& 42& 84\%& 40& 80\%& 40& 80\%\\
Fairness & 41& 82\%& 36& 72\%& 41& 82\%& 39& 78\%\\
Robustness & 41& 82\%& 41& 82\%& 40& 80\%& 39& 78\%\\
\hline
\multicolumn{9}{|c|}{\textbf{QA - Standard Few shot}} \\
\hline
Taxonomy & 41& 82\%& 37& 74\%& 39& 78\%& 34& 68\%\\
Negation & 41& 82\%& 30& 60\%& 39& 78\%& 31& 62\%\\
Coreference & 41& 82\%& 41& 82\%& 39& 78\%& 42& 84\%\\
SRL & 42& 84\%& 42& 84\%& 38& 76\%& 40& 80\%\\
Vocab & 41& 82\%& 44& 88\%& 39& 78\%& 39& 78\%\\
Fairness & 42& 84\%& 39& 78\%& 38& 76\%& 39& 78\%\\
Robustness & 42& 84\%& 43& 86\%& 41& 82\%& 39& 78\%\\
\hline
\end{tabular}
}
\label{tab:qna_performance}
\end{table}

\begin{table}[t]
\centering
\caption{QA (with context) Results: GPT-4o vs Gemini-2.0-Flash}
\small
\resizebox{\textwidth}{!}{
\begin{tabular}{|l|cc|cc|cc|cc|}
\hline
\multirow{3}{*}{Category} & \multicolumn{4}{c|}{\textbf{GPT-4o}} & \multicolumn{4}{c|}{\textbf{Gemini-2.0-Flash}} \\
\cline{2-9}
& \multicolumn{2}{c|}{Accuracy} & \multicolumn{2}{c|}{{\sf MMT4NL}} & \multicolumn{2}{c|}{Accuracy} & \multicolumn{2}{c|}{{\sf MMT4NL}
} \\
& \#Correct& Pass Rate & \#Correct& Pass Rate & \#Correct& Pass Rate & \#Correct& Pass Rate \\
\hline
\multicolumn{9}{|c|}{\textbf{QA - Standard zero shot}} \\
\hline
Taxonomy & 47 & 94\%& 47 & 94\%& 47 & 94\%& 46 & 92\%\\
Negation & 47 & 94\%& 42 & 84\%& 47 & 94\%& 32 & 64\%\\
Coreference & 47 & 94\%
& 49 & 98\%& 48 & 96\%& 48 & 96\%\\
SRL & 47 & 94\%& 48 & 96\%& 47 & 94\%& 46 & 92\%\\
Vocab & 47 & 94\%
& 47 & 94\%& 47 & 94\%& 46 & 92\%\\
Fairness & 47 & 94\%
& 45 & 90\%& 48 & 96\%& 43 & 86\%\\
Robustness & 47 & 94\%& 47 & 94\%& 46 & 92\%& 46 & 92\%\\
\hline
\multicolumn{9}{|c|}{\textbf{QA - Standard One shot}} \\
\hline
Taxonomy & 48 & 96\%& 47 & 94\%& 47 & 94\%& 46 & 92\%\\
Negation & 49 & 98\%& 41 & 82\%& 47 & 94\%& 37 & 74\%\\
Coreference & 48 & 96\%& 49 & 98\%& 49 & 98\%& 47 & 94\%\\
SRL & 49 & 98\%& 49 & 98\%& 47 & 94\%& 47 & 94\%\\
Vocab & 48 & 96\%& 48 & 96\%& 47 & 94\%& 46 & 92\%\\
Fairness & 48 & 96\%& 45 & 90\%& 48 & 96\%& 45 & 90\%\\
Robustness & 48 & 96\%& 49 & 98\%& 47 & 94\%& 48 & 96\%\\
\hline
\multicolumn{9}{|c|}{\textbf{QA - Standard Few shot}} \\
\hline
Taxonomy & 48 & 96\%& 47 & 94\%& 49 & 98\%& 47 & 94\%\\
Negation & 48 & 96\%& 39 & 78\%& 47 & 94\%& 32 & 64\%\\
Coreference & 48 & 96\%& 48 & 96\%& 49 & 98\%& 48 & 96\%\\
SRL & 48 & 96\%& 49 & 98\%& 48 & 96\%& 46 & 92\%\\
Vocab & 48 & 96\%& 47 & 94\%& 48 & 96\%& 48 & 96\%\\
Fairness & 48 & 96\%& 47 & 94\%& 49 & 98\%& 45 & 90\%\\
Robustness & 48 & 96\%& 48 & 96\%& 48 & 96\%& 47 & 94\%\\
\hline
\end{tabular}
}
\label{tab:qna_context_performance}
\end{table}

\subsection{Question Answering}

\subsubsection{QA with no Context:}

Table \ref{tab:qna_performance} shows that without question context, GPT-4o’s accuracy ranged from 78\% to 82\%, and Gemini-2.0-Flash’s from 72\% to 88\%. Similar to the sentiment analysis task, {\sf MMT4NL} effectively revealed interpretation flaws, as both models gave different answers to perturbed inputs, with pass rates varying from 60\% to 88\%. One-shot and few-shot prompts didn’t significantly improve performance, except for Gemini-2.0-Flash, which showed better pass rates for Fairness and Robustness.

\subsubsection{QA with Context:}

Providing context improved both LLMs' performance (see Table~\ref{tab:qna_context_performance}). Accuracy increased to 94\%–98\% across all perturbations. GPT-4o's pass rate rose to 78\%–98\%, while Gemini-2.0-Flash's increased to 64\%–96\%. Interestingly, the one-shot and few-shot prompt templates did not clearly show effectiveness in improving the pass rates for both models.

\subsection{Observations}
We manually investigated incorrect results and only showed surprising and severe scenarios in this paper (due to space limitation) as follows.

\subsubsection{Fairness:}

Gendered perturbations could alter sentiment classification, revealing potential bias. For example, ``The traffic is heavy'' (Negative) became Neutral when changed to ``She thinks the traffic is heavy.'' Similarly, ``The traffic is light'' (Positive) became Neutral when changed to ``A female thinks the traffic is light''. This suggests possible gender stereotyping in traffic-related opinions. 
A similar observation was found in question-answering tasks. GPT-4o answered No for ``Would it be impossible to seat every Chief Justice of the United States on a Boeing 737?'' but answered Yes if changing ``Chief Justice'' to ``\textbf{female} Chief Justice''.
In contrast, adding nationality or ethnicity modifiers (e.g., ``dress'' $\rightarrow$ ``Indian dress'') did not affect sentiment classification.

\subsubsection{Taxonomy:} 
Models show varying sensitivity to different types of word substitutions, revealing inconsistencies in sentiment interpretation.

\begin{itemize}
    \item \textbf{Feelings Replacement:} LLMs could struggle with intensity-modifying synonyms, shifting sentiment despite preserving core meaning. GPT-4o had a 28.6\% failure rate (2/7) %
    and Gemini-2.0-Flash 37.5\% (3/8). For example, GPT-4o correctly classified ``I’m really hungry'' as Neutral but misclassified ``I’m really starving'' as Negative. This highlights how stronger emotional intensity can distort sentiment classification.

    \item \textbf{Action Verb Replacement:} This caused the highest failure rates — 71.4\% for GPT-4o (5/7) and 62.5\% for Gemini-2.0-Flash (5/8). For example, ``This software is very complicated'' (Negative) was misclassified as Neutral when changed to ``This software is very complex.'' This suggests that models anchor sentiment inconsistently to action verbs, even when meanings are similar.
    A similar observation was found in question answering for GPT-4o. ``Does Orange County, California require airplanes to be quiet?'' resulted in Yes while changing ``quiet'' to ``silent?'' resulted in No.

    \item \textbf{Descriptive Word Replacement:} Models handled descriptive synonyms well, showing no failures when replacing terms like ``poor'' $\rightarrow$ ``bad'' or ``stunning'' $\rightarrow$ ``gorgeous.'' This indicates stronger and more stable sentiment associations with adjectives and descriptive terms.

    \item \textbf{Generalization to Broader Categories:} Models maintained consistent sentiment when specific terms were replaced with broader categories (e.g., ``fitness'' $\rightarrow$ ``exercise''), showing resilience to category-level abstraction.
    
\end{itemize}

\subsubsection{Robustness:}This simple perturbation, i.e., swapping characters, also revealed some flaws in the models. For example, Gemini-2.0-Flash answered this question
``Would the United States Military Academy reject an applicant with multiple sclerosis?'' with Yes. However, after changing ``sclerosis'' to ``scle\textbf{or}sis'', the answer changed to No. The failure may stem from tokenization processes that treat these altered spellings as entirely different tokens, changing the semantic representation significantly. This vulnerability reveals limitations in models' ability to recognize misspelled medical or technical terms when the misspelling creates a non-existent word, as opposed to common misspellings that might appear in training data.

\subsubsection{Coreference:}Restructuring questions to include explicit pronoun references creates referential distance between the subject and its attributes, challenging the model's ability to maintain logical connections. For example, ``Can an anchovy born in 2020 survive 25th US census?'' was answered with a No by GPT-4o. Nonetheless, ``\textbf{Considering an anchovy born in 2020,} can it survive the 25th US census?'' was answered with a Yes.

\section{Conclusion}

We presented {\sf NNT4ML}, a unified framework for testing the trustworthiness 
degree of LLMs, especially on the ICL capabilities. 
It clearly combines two strong advantages of CheckList (i.e. linguistic interpretability) and PromptBench (i.e. an adversarial attack framework on LLMs). Our findings 
reveal surprising bugs in state-of-the-art LLMs, 
indicating that it complements current practices (exact match) well. It is expected that tests created according to {\sf NNT4ML} 
can be applied for any prompt construction. 
However, the framework’s
effectiveness depends on the quality and diversity of transformation functions (perturbations). Addressing
these limitations is a priority for our ongoing and future work.

\bibliographystyle{splncs04}
\bibliography{mybibliography}

\end{document}